\documentclass[12pt,twoside]{article}
\usepackage[mathscr]{eucal}
\usepackage{amsmath,amsfonts,amssymb,amsthm,mathabx,empheq}
\usepackage{times}
\usepackage{pdfsync}
\usepackage{cite}
\usepackage{url}
\usepackage{hyperref}
\usepackage{tensor}
\usepackage{color}
\usepackage{multicol}
\usepackage{bbold}

\advance\textheight\topskip
\allowdisplaybreaks[1]

\newcommand\td{\text{d}}
\newcommand\cO{{\cal O}}
\newcommand{\p}{\partial}

\newcommand{\be}{\begin{equation}}
\newcommand{\ee}{\end{equation}}
\newcommand{\bea}{\begin{eqnarray}}
\newcommand{\eea}{\end{eqnarray}}

\def\bz{\bar z}

\def\n{\nabla}

\def\nb{\bar\nabla}

\newcommand{\nn}{\nonumber}
\newcommand*\xbar[1]{%
  \hbox{%
    \vbox{%
      \hrule height 0.5pt 
      \kern0.3ex
      \hbox{%
        \kern-0.0em
        \ensuremath{#1}%
        \kern-0.0em
      }%
    }%
  }%
}

\DeclareFontFamily{OT1}{rsfs}{} \DeclareFontShape{OT1}{rsfs}{m}{n}{
<-7> rsfs5 <7-10> rsfs7 <10-> rsfs10}{}
\DeclareMathAlphabet{\mycal}{OT1}{rsfs}{m}{n}

\hypersetup{
colorlinks=true,
linktoc=page,    
linkcolor=blue,
citecolor=blue,
urlcolor=blue,
colorlinks=true,
}

\begin{document}

\title{Supertranslation ambiguity in post-Minkowskian expansion}

\author{Pujian Mao and Baijun Zeng}
\date{}

\def\mytitle{Supertranslation ambiguity in post-Minkowskian expansion}

\addtolength{\headsep}{4pt}

\begin{centering}

  \vspace{1cm}

  \textbf{\large{\mytitle}}

  \vspace{1cm}

  {\large Pujian Mao and Baijun Zeng}

\vspace{0.5cm}

\begin{minipage}{.9\textwidth}\small \it  \begin{center}
     Center for Joint Quantum Studies and Department of Physics,\\
     School of Science, Tianjin University, 135 Yaguan Road, Tianjin 300350, China
 \end{center}
\end{minipage}

\end{centering}

\begin{center}
Emails:  pjmao@tju.edu.cn,\,\,zeng@tju.edu.cn
\end{center}

\begin{center}
\begin{minipage}{.9\textwidth}
\textsc{Abstract}: The supertranslation ambiguity of angular momentum is a long-standing problem in general relativity,  which arises the puzzle of the angular momentum loss in the post-Minkowskian (PM) expansion. The angular momentum loss in gravitational scattering is of order $G^2$ in the linear response theory but of order $G^3$ in the amplitudes-based methods. In this paper, we propose a generic prescription to fix the supertranslation ambiguity in the PM expansion which will uniquely determine the angular momentum loss. The proposal is self-contained and involves only the null infinity data.

\end{minipage}
\end{center}

\thispagestyle{empty}


\section{Introduction}

The success of direct observation of gravitational waves opens new avenues for gravitational wave astronomy \cite{LIGOScientific:2016aoc}. The accurate theoretical modeling of gravitational scattering is crucial for the gravitational wave detectors \cite{Buonanno:2022pgc}. Currently, the waveform of gravitational radiation is mainly derived in expansion based on successive approximations in powers of some small parameters, such as the inverse of the speed of light $1/c$ or the Newton's constant $G$, namely the post-Newtonian or post-Minkowskian expansions, respectively, see, e.g., the reviews in \cite{Blanchet:2013haa,Bjerrum-Bohr:2022blt}. A common feature of those approximate methods is that the Einstein equation is represented by a linear wave equation at each order in the expansion on the Minkowski or curved background and the non-linear effect of the Einstein theory is encoded in the effective stress tensor term, which raises huge disputes that if the gravitational radiation was just an artifact of the linearization from the expansion up until the 1960s. The dispute is settled by Bondi and collaborators by formulating the Einstein equation as a characteristic initial value problem at null infinity \cite{Bondi:1962px,Sachs:1962wk}. In the Bondi's framework, the gravitational radiation is characterized by the news functions in the asymptotic metric and the mass of the system always decreases whenever news functions exist, which eventually confirms the existence of gravitational waves in the full Einstein theory.

A surprising result from the Bondi's framework \cite{Bondi:1962px,Sachs:1962wk} is that the asymptotic symmetry group at the null infinity is not the Poincar\'{e} group, but the semidirect sum of supertranslations, an infinite dimensional Abelian group which extends the translations, with the global conformal group of the celestial sphere that is isomorphic to the Lorentz group. The enhancement from translation to supertranslation has several remarkable applications recently, see, e.g., the comprehensive review in \cite{Strominger:2017zoo}, but it also brings puzzles in general relativity. A long-standing problem is the supertranslation ambiguity in defining angular momentum \cite{Penrose,Held}. In the context of the PM expansion, the supertranslation ambiguity causes the puzzle of angular momentum loss in gravitational scattering \cite{Veneziano:2022zwh}. The lowest order of the angular momentum loss can be at $\cO(G^2)$ or at $\cO(G^3)$ depending on different gauge choices under the supertranslation \cite{Damour:2020tta,Jakobsen:2021smu,Mougiakakos:2021ckm,Herrmann:2021lqe,Herrmann:2021tct,Bini:2021gat,Riva:2021vnj,Veneziano:2022zwh,Manohar:2022dea,DiVecchia:2022owy,Bini:2022enm,Bini:2022wrq,Heissenberg:2022tsn,Heissenberg:2023uvo,Heissenberg:2024umh}. The angular momentum loss formula at null infinity and its relation to the change of angular momentum of the source are certainly of utmost importance for higher precision PM computations, which leads to the intensive investigations recently from various perspectives, including the detailed analysis of the compact binary coalescence systems \cite{Ashtekar:2019rpv,He:2022idx}, supertranslation invariant definitions of the angular momentum at null infinity \cite{Compere:2019gft,Chen:2021szm,Compere:2021inq,Chen:2021kug,Javadinezhad:2022hhl,Chen:2022fbu,Javadinezhad:2022ldc} and at spatial infinity \cite{Fuentealba:2022xsz,Fuentealba:2023syb}, supertranslation frame-fixing from the matching condition \cite{Veneziano:2022zwh,Ashtekar:2023zul} and from covariance \cite{Javadinezhad:2023mtp}, memory effect \cite{Mao:2023evc,Riva:2023xxm}, and flux balance law \cite{Riva:2023xxm,Gralla:2024wzr}.

In this paper, we propose to use asymptotic charges in the PM expansion to fix the supertranslation ambiguity. Our proposal is self-contained at the null infinity and does not involve any data from the spatial infinity. The resolution is encoded in the linear nature of the wave equation in the PM expansion. When the Einstein equation is expanded about the Newton's constant $G$, one can recover a linear wave equation at each order of $G$ which is sourced by the stress tensor from the matter fields at each order and the effective stress tensor consisting of the perturbative metric in the lower orders of $G$ that reflects the non-linear feature of the Einstein theory. At each order of $G$, the sourced wave equation is invariant under gauge transformations which are inherited from the diffeomorphism invariance of the Einstein theory. In particular, there is a supertranslation at each order of $G$ when combining the PM expansion with the Bondi's framework \cite{Compere:2019odm}.

Consider that the full metric is expanded around the background Minkowski spacetime to the second order in $G$ as $g_{\mu\nu}=\eta_{\mu\nu}  +  G H_{\mu\nu} + G^2 h_{\mu\nu}$. Correspondingly, a diffeomorphism transformation in the infinitesimal version can be also expanded as
\be
\delta g_{\mu\nu}\equiv {\cal L}_\xi g_{\mu\nu}={\cal L}_\xi (\eta_{\mu\nu} + G H_{\mu\nu} + G^2 h_{\mu\nu}).
\ee
If one also expands the symmetry parameter in $G$ as $\xi=G \xi^{(1)} + G^2 \xi^{(2)}$,\footnote{One should consider $G$ as a dimensionless parameter. Otherwise, it should be $G M_c/(c^2 r_c)$, where $M_c$ is a mass parameter and $r_c$ is a radial parameter. Here, we introduce the $G$ expansion in the symmetry transformation as an effective way of deriving gauge transformation of the theory at each order of $G$, see also \cite{Veneziano:2022zwh} for a similar treatment.} the induced infinitesimal gauge transformation at order $G$ is
\be
\delta H_{\mu\nu} = {\cal L}_{\xi^{(1)}} \eta_{\mu\nu},
\ee
which is a pure gauge transformation that will not change the curvature tensor at order $G$. Correspondingly, the supertranslation at order $G$ will not change the leading order of the linear and angular momentum in the expansion of $G$ \cite{Mao:2024urq}.\footnote{The
  supertranslation at order $G$ will affect the angular momentum loss at the
  leading order \cite{Veneziano:2022zwh}. But the leading order angular momentum, which is different from the leading order angular momentum loss, is independent of the order $G$ supertranslation.} At order $G^2$, the equations of motion are invariant under the transformation
\be\label{2ndorder}
\delta h_{\mu\nu} = {\cal L}_{\xi^{(2)}} \eta_{\mu\nu} + {\cal L}_{\xi^{(1)}} H_{\mu\nu}.
\ee
The first piece on the right hand side is the standard gauge transformation. The second piece is relevant to the gauge transformation at order $G$, which is not necessary to be a pure gauge transformation at order $G^2$. The perturbative metric at order $G^2$ satisfies the Fierz-Pauli equation sourced by two parts, the stress tensor of the matter fields $T_{\mu\nu}^M$ with corrections from the perturbative metric at order $G$ and the effective stress tensor $T_{\mu\nu}^E$ constructed purely from the order $G$ metric,
\begin{align}
E_{\mu\nu}\equiv & \left[\nb_\tau \nb_\mu h ^\tau_\nu + \nb_\tau \nb_\nu h ^\tau_\mu - \nb^2 h_{\mu\nu} - \nb_\mu \nb_\nu h\right]-  \eta_{\mu\nu}\left(\nb_\alpha \nb_\beta h^{\alpha\beta} - \nb^2 h\right) \nn \\
 =& T_{\mu\nu}^E+T_{\mu\nu}^M,
\end{align}
where $h=\eta^{\mu\nu}h_{\mu\nu}$, $H=\eta^{\mu\nu}H_{\mu\nu}$, and we use $\nb$ to denote the derivative in the Minkowski spacetime. The Fierz-Pauli equation is linear, which guarantees that the metric can be a superimposition as $h_{\mu\nu}=h_{\mu\nu}^S+h_{\mu\nu}^M$ where the first piece $h_{\mu\nu}^S$, which we would refer to as the
self-part,\footnote{The self-part in our terminology includes
  the self-terms and cross-terms in a two pointlike objects system investigated
  in \cite{Bel:1981be}. The split of the metric can be obviously performed in the Cartesian coordinates system. The equations of motion for both $h^S_{\mu\nu}$ and $h^M_{\mu\nu}$ are in a covariant way. Hence, one can write both equations in any other coordinates system.} satisfies the wave equation sourced by $T_{\mu\nu}^E$ and the second piece $h_{\mu\nu}^M$ satisfies the wave equation sourced by $T_{\mu\nu}^M$. One can verify that both $T_{\mu\nu}^E$ and $T_{\mu\nu}^M$ are conserved. Hence, the equations for both $h_{\mu\nu}^S$ and $h_{\mu\nu}^M$ are invariant under the transformation represented by the first piece on the right hand of \eqref{2ndorder}. However, the second piece on the right hand of \eqref{2ndorder} is not an independent gauge transformation for both $h_{\mu\nu}^S$ and $h_{\mu\nu}^M$.

Considering only the self-part, we will show that the next-to-leading order angular momentum in the expansion of $G$ is mainly controlled by the order $G$ supertranslation. Thus, the prescription we will propose to fix the order $G$ supertranslation, hence to resolve the supertranslation ambiguity of the angular momentum loss in the PM expansion, is from a physical requirement of the next-to-leading order angular momentum from the self-part. In Einstein theory, the supertranslation freedom is intrinsic in the sense that it is associated with the propagating degree of freedom. In other words, fixing the gauge choice associated to the supertranslation freedom is equivalent to turning off the gravitational radiation \cite{Held}, see also \cite{Strominger:2014pwa} for a modern interpretation. It is well known that the propagating degree of freedom represented by the news functions starts at order $G^2$ in the PM expansion, see, e.g., \cite{Veneziano:2022zwh,Damour:2020tta}. Thus we can fix the supertranslation at order $G$ without touching the propagating degree of freedom at order $G^2$, which is a fine tuning structure arising in the PM expansion. The key point now is what the physical requirement of the next-to-leading order angular momentum for the self-part should be.


\section{The resolution to supertranslation ambiguity}

The criteria we propose to fix the order $G$ supertranslation is that the self-part $h_{\mu\nu}^S$ does not contribute to the next-to-leading order linear and angular momentum. The effective stress tensor at order $G^2$ is given by \cite{Bel:1981be,MTW}
\begin{align}
   T^E_{\mu\nu}=&  \eta_{\mu\nu} \bigg[  H^{\alpha\beta} \left(\nb_\alpha \nb_\beta H + \nb^2 H_{\alpha\beta} - 2 \nb^\tau \nb_\alpha H_{\tau\beta} \right) -\frac{1}{2}\nb_{\beta}H_{\alpha\tau}\nb^{\tau}H^{\alpha\beta}  \nn \\
  &    +  \nb_\beta H \left(\nb_\alpha H^{\alpha \beta} - \frac14 \nb^\beta H\right)  - \nb_\alpha H^{\alpha\beta} \nb^\tau H_{\tau \beta} + \frac34 \nb_\alpha H^{\tau\beta} \nb^\alpha H_{\tau^\beta} \bigg]\nn \\
  &+  \nb_\alpha H^{\alpha\tau} \left(\nb_\mu H_{\nu\tau} + \nb_\nu H_{\mu\tau} - \nb_\tau H_{\mu\nu}\right)\nn \\
  & + H^{\tau\alpha} \big(\nb_\alpha \nb_\mu H_{\nu\tau} + \nb_\alpha \nb_\nu H_{\mu\tau } - \nb_\alpha \nb_\tau H_{\mu\nu} - \nb_\mu \nb_\nu H_{\tau\alpha} \big)  \nn \\
  & 
 +  H_{\mu\nu} \left(\nb_\alpha \nb_\beta H^{\alpha\beta} - \nb^2 H \right)  - \nb_\alpha H_{\nu\tau} \nb^\alpha H_\mu^\tau
 - \frac12 \nb_\mu H_{\tau\alpha} \nb_\nu H^{\tau\alpha}\nn \\
& +   \nb_\alpha H_{\nu\tau} \nb^\tau H_\mu^\alpha   -\frac12 \nb^\alpha H \left(\nb_\mu H_{\nu\alpha} + \nb_\nu H_{\mu\alpha} - \nb_\alpha H_{\mu\nu} \right).
\end{align}
The order $G$ metric $H_{\mu\nu}$ of the spacetime sourced by a system of pointlike bodies is known and transformed into the Bondi's framework recently \cite{Mao:2024urq}. Up to relevant orders in $1/r$ series expansion near null infinity, the metric is given by
\begin{align}
& H_{uu}=  \frac{2 M_b }{r} +  \cO(\frac{1}{r^{2}}),\quad \quad M_b= - \sum{\frac{m}{(v_{\mu}n^{\mu})^{3}}}, \\
& H_{ur}=  \cO(\frac{1}{r^{2}}),\quad \quad H_{AB}= r  C_{AB} +  \cO(1), \\
 & C_{AB}=\left[ 2D_{{A}}D_{{B}} - {\gamma}_{{A}{B}} D_{{C}}D^{{C}} \right] \left[\bar\beta(x^A) - f(x^A)\right],\label{CAB} \\
 &\quad\quad f(x^A)= 2 \sum m (n\cdot v) \log(-(n\cdot v)),\nn\\
& H_{uA}=\frac{1}{2} D^B C_{AB} +\frac{2}{3r}\left[u D_{A}\left( M_b \right) +  N_A  \right]+\cO(\frac{1}{r^2}),\\
 &N_A=D_{A}\bigg\{\frac{3}{2}\sum m\bigg[-\frac{(n^{\mu}c_{\mu})^{2}}{(n^{\nu}v_{\nu})^{3}}+\frac{v^{\mu}c_{\mu}n^{\nu}c_{\nu}}{(n^{\sigma}v_{\sigma})^{2}}  +\frac{v^{\mu}c_{\mu}}{(n^{\nu}v_{\nu})^{2}} \nn\\
 &\quad\quad +\frac{1}{3}\frac{n^{\mu}c_{\mu}}{(n^{\nu}v_{\nu})^3} \bigg]  \bigg\}  -\sum{m\frac{2}{(n^{\nu}v_{\nu})^{3}}D_{A}(n^{\mu}c_{\mu})} ,
\end{align}
where $\gamma_{AB}$ is the boundary metric of the celestial sphere and $D_A$ is the covariant derivative with respect to $\gamma_{AB}$. The capital index will be lowered and raised by $\gamma_{AB}$ and its inverse metric. Here, $\bar\beta(x^A)$ is an arbitrary function on the celestial sphere that characterizes the order $G$ supertranslation. $v^\mu$ and $c^\mu$ are the 4-velocities and initial positions of the particle. $n^\mu=(1,n^i)$ in the Cartesian coordinates where $n^i$ is the normal vector defined from the relation to the spatial parts of the Cartesian coordinates $x^i=r n^i$. The sign $\sum$ denotes the sum of contributions from all particles. For other notations, we would refer to \cite{Mao:2024urq}. Inserting the order $G$ metric into the effective stress tensor, we obtain, up to relevant order in $1/r$ expansion,
\begin{align}
&T^E_{rr}=-\frac{C^{AB}C_{AB}}{2r^4} + \cO(r^{-5}),\quad T^E_{ur}=\cO(r^{-4}),\nn\\
&  T^E_{AB}=\frac{\gamma_{AB} C^{EF}N_{EF}}{2r} + \cO(r^{-2}), \quad N_{AB}=\p_u C_{AB},\nn\\
&T^E_{rA}=\frac{D_A (C^{AB}C_{AB})}{4r^3} +\cO(r^{-4}),\\
&T^E_{uA}=- \frac{D^B C_{AE} {N_B}^E + 3 C_{AB} D_E N^{BE}}{2r^2} + \cO(r^{-3}), \nn\\
& T^E_{uu}= - \frac{N^{AB} N_{AB} + 2 C_{AB} \p_u N^{AB}}{2r^2} + \cO(r^{-3}).\nn
\end{align}
It is clear that $N_{AB}=0$ from the expression in \eqref{CAB}, which simply confirms that there is no news function at order $G$. The components $T^E_{uu}$ and $T^E_{uA}$ from the effective stress tensor decay faster than the case of a pointlike particle at large distance, see, e.g., the asymptotic behaviors of the stress tensor in the Appendix of \cite{Pasterski:2015tva}. As will be detailed in the next pages, those components of the effective stress tensor control the evolution of the asymptotic charges from the self-part, the absence of which means that there is no source driving the next-to-leading order charges of the self-part. In the standard asymptotic analysis \cite{Bondi:1962px,Sachs:1962wk}, it is normally dealing with vacuum solution space, which only means that there is no local source closed to the null infinity. There can be matter fields inside the bulk of the spacetime. The non-trivial asymptotic charges simply indicate the information of the matter fields in the bulk. Since the effective stress tensor is defined locally, the linear and angular momentum from the self-part at the next-to-leading order should vanish with respect to the asymptotic behaviors of the effective stress tensor, which provides the main justification of our prescription. 

We will prove that the order $G$ supertranslation freedom can be completely fixed by our prescription. We apply the asymptotic analysis to derive the generic order $G^2$ solutions with the effective stress tensor in series expansion near null infinity, where the initial data of the order $G^2$ metric will be fixed by the criteria of vanishing of the linear and angular momentum. We then compute the transformation laws of the linear and angular momentum under the order $G$ supertranslation. We show that the order $G$ supertranslation does not preserve the next-to-leading order angular momentum, hence, should be fixed by the vanishing condition of the angular momentum.

A curious point is that 
the supertranslation is only defined at null infinity but the PM system can be only solved in the bulk without touching any infinity data. The supertranslation dependence of the angular momentum loss from the bulk point of view can be better understood from the recently established amplitudes-based methods. The angular momentum loss at order $G^2$ is only relevant to the zero-frequency gravitons, see, e.g., \cite{Manohar:2022dea,DiVecchia:2022owy}, which have clear connections to the null infinity supertranslation from the triangle equivalent relations \cite{Strominger:2013jfa,He:2014laa}, see also \cite{Strominger:2017zoo}.


\section{Solution space at order $G^2$ and asymptotic charges}

We will solve the order $G^2$ self-part theory in series expansion near future null infinity. All of the metric components are assumed to be given in $1/r$ series expansion. For notational brevity, we suppress the label $S$ in the metric. We will adapt the Bondi gauge to the order $G^2$ theory where it is easy to read the following conditions $h_{rr}=h_{rA}=0$. The determinant condition in the Bondi gauge will induce constraints between the trace of $h_{AB}$ and the data at $\cO(G)$. The precise condition is
\be\label{determinant}
 h_{AB}=r c_{AB} + \frac14  \gamma_{AB} C_{EF} C^{EF} + d_{AB} + \cO(r^{-1}),
\ee
where $c_{AB}$ and $d_{AB}$ are traceless. This condition seems artificial. But the determinant condition indeed mixes data between different orders of the expansion in $G$. One can alternatively use the Newman-Unti (NU) gauge \cite{Newman:1962cia} (see also the application in linearized gravity \cite{Conde:2016rom}) where one can set the gauge condition $h_{ur}=0$. We have checked that transforming the solution in the NU gauge to the Bondi gauge will fix the subleading order of $ h_{AB}$ as \eqref{determinant}. The boundary conditions adapted from the Bondi gauge are $h_{ur}=\cO(r^{-2})$, $h_{uA}= \cO(1)$, and $h_{uu}=\cO(r^{-1})$. The effective stress tensor is conserved outside the trajectories of the pointlike bodies. Thus, we can use the well-known organization of the Einstein equation in the Bondi gauge \cite{Bondi:1962px,Sachs:1962wk,Barnich:2010eb,Conde:2016rom} to solve the order $G^2$ theory near null infinity. The four hypersurface equations $E_{r\alpha}=T^E_{r\alpha}$ will determine $h_{ur}$, $h_{uA}$, and $h_{uu}$. In $1/r$ series expansion, we obtain
\begin{align}
 & h_{ur}=\frac{1}{16 r^2} C_{AB} C^{AB}  + \cO(r^{-3}),\\
 &h_{uA}=D^B c_{AB}+ \frac{1}{r}\bigg[ \frac23 (n_A + u D_A m_b) 
  - \frac{1}{16} D_A (C_{AB} C^{AB}) \bigg]  + \cO(r^{-2}),  \\
 & D^B d_{AB}=0, \quad\quad h_{uu}=  \frac{2 m_b}{r} + \cO(r^{-2}),
\end{align}
where $m_b$ and $n_A$ are integration constants from the differential equations of $r$. Two standard equations, the traceless part of $E_{AB}=T^E_{AB}$, will fix the evolution of the data at each order of $h_{AB}$ in the $1/r$ expansion except $c_{AB}$. In particular, we have $\p_u d_{AB}=0$ from the leading order. The trace part of $E_{AB}=T^E_{AB}$ is trivial when the previous six equations of motion are satisfied. Three supplementary equations will fix the time evolution of the integration constants as
\begin{multline}
\p_u n_A = \frac14 D^B D_A D^E c_{EB} -\frac14 D^2 D^B c_{BA} - u D_A (\p_u m_b)\\
+\frac14 D_A (C_{BE}N^{BE}) - \frac14 D_B (C^{BE}N_{EA}) + \frac12 C_{AB} D_E N^{EB},
\end{multline}
and
\be
\p_u m_b=\frac14 D_A D_B n^{AB} - \frac18 N_{AB} N^{AB},
\ee
where $n_{AB}=\p_u c_{AB}$. The results are consistent with the solution space of full Einstein theory near null infinity \cite{Barnich:2010eb,Flanagan:2015pxa}. Inserting the precise formula of $C_{AB}$ from the previous section, the evolution relations are reduced to
\begin{align}
&\p_u n_A = \frac14 D^B D_A D^E c_{EB}
-\frac14 D^2 D^B c_{BA}- \frac14 u D_A D_B D_E n^{BE},\label{nAu}\\
&\p_u m_b=\frac14 D_A D_B n^{AB}.\label{mbu}
\end{align}

The Bondi linear and angular momentum at the leading order in the expansion of $G$ coincide with the Arnowitt–Deser–Misner (ADM) definitions \cite{Mao:2024urq},
\be
P^\mu_{\text{ADM}}=\sum p^\mu,\quad \quad J^i_{\text{ADM}}={\epsilon^i}_{jk} \sum c^j p^k.
\ee
The next-to-leading order contributions of the Bondi linear and angular momentum at null infinity are \cite{Barnich:2011mi,Flanagan:2015pxa}
\be
\begin{split}\label{charge}
&P^{\mu}=\frac{G}{4\pi }\int_{S} \, m_b n^{\mu} \, \td\Omega,\\
&J^{(i)}=\frac{G}{8\pi }\int_{S} \, Y^{(i) A} \left(n_{ A} -\frac14 C_{AB} D_E C^{BE} \right)\, d\Omega,
\end{split}
\ee
where $\td \Omega$ is the integral measure on the celestial sphere and $ Y^{(i) A} \frac{\p}{\p x^{A}}$ are three Killing vectors of the celestial sphere, which are related to the normal vector $n^i$ by $Y^{(i) A}=-\epsilon^{ A  B}D_{ B}n^{i}$. 

The vanishing of the next-to-leading order Bondi linear and angular momentum will fix $m_b$ and $n_A$ as
\begin{align}
&m_b=\frac14 D_A D_B c^{AB}+......,\label{m}\\
&n_A = \frac14 \int \, \td u \, \bigg(D^B D_A D^E   c_{EB}  - D^2 D^B  c_{BA}  -  u D_A D_B D_E n^{BE}  \bigg)+ ...... ,\label{nA}
\end{align}
where the ellipses denote $u$-independent total derivative terms which can not be fixed by the charge analysis and we have applied the identity \cite{Compere:2016jwb,Javadinezhad:2022hhl}
\begin{multline}
-\frac14 Y^{(i) A} C_{AB} D_E C^{BE}=
\frac14 D^A \bigg[Y^{(i) B} C_{AB}(D^2+2) 
\left[ f(x^A) - \bar\beta(x^A) \right] \\
+\frac12 Y^{(i)}_A \big((D^2+2)  \left[ f(x^A) - \bar\beta(x^A) \right] \big)^2\bigg],
\end{multline}
according to the expression of $C_{AB}$ in \eqref{CAB}. 
Though we have fixed $m_b$ and $n_A$ from the asymptotic charges, the relevance to the order $G$ supertranslation is not yet uncovered. We will compute the transformation law of $m_b$ and $n_A$ under the order $G$ supertranslation and prove that a generic order $G$ supertranslation can not preserve the form of $n_A$ in \eqref{nA}.


\section{Transformation laws under the order $G$ supertranslation}

An order $G$ supertranslation connecting two solutions in the PM expansion up to order $G^2$ in the Bondi gauge can be effectively obtained by the coordinates transformation from $(u',r',{x'}^A)$ to $(u,r,x^A)$. In $1/r$ expansion, we obtain up to relevant orders as
\begin{align}
& {u'}=u - G T(x^{A}) - \frac{G^2}{2r} D^A T D_A T + \cO(G^3), \\
&r'=r-\frac{G}{2}D^2T+\frac{G^2}{4r}\bigg(D^{A}D^{B}TD_{A}D_{B}T + 2 D_{A}C^{AB}D_{B}T +C^{AB}D_{A}D_{B}T\nn\\
    &\quad \quad  +2D_{A}TD^{A}T-\frac{1}{2}D^2TD^2T\bigg)+\frac{G^2}{6r^2}D^{AB}D_{A}D_{B}T+\cO(G^3)\\
&{{x'}}^{A}=x^A+\frac{G}{r}D^{A}T+\frac{G^2}{2r^2}\bigg( D_{B}TD^{A}D^{B}T-C^{AB}D_{B}T +D^{A}TD^{2}T \nonumber\\
&\quad   -\gamma^{AB}\partial_{B}D^{C}TD_{C}T  +\partial_{C}\gamma^{AB}D_{B}TD^{C}T \bigg)  -\frac{G^2}{3r^3}D^{AB}D_{B}T + \cO(G^3) ,
\end{align}
where $T(x^{A})$ represents a supertranslation at order $G$, $\gamma=\sqrt{\det{\gamma_{AB}}}$, and $D_{AB}$ is the subleading order term of $H_{AB}$ in the $1/r$ expansion. We will not give the precise form of $D_{AB}$ as it will not affect the transformation law. Since our goal is to fix the order $G$ supertranslation, we did not include the order $G^2$ supertranslation. For computational simplicity, we choose the complex stereographic coordinates $A=(z,\bz)$ on the celestial sphere. The transformation laws of the Bondi mass aspect, angular momentum aspect, and asymptotic shear at order $G$ are obtained as
\be
\begin{split}
&M_b= {M'}_b, \quad \quad  N_A = {N'}_A, \\
&C_{AB}= {C'}_{AB} + (2D_A D_B - \gamma_{AB}D^2) T,
\end{split}
\ee
where the primed quantities are functions in the old coordinates system. The results are in accordance with the fact that the leading order Bondi linear and angular momentum are irrelevant to the order $G$ supertranslation \cite{Mao:2024urq}. The transformation laws at order $G^2$ are
\be
m_b= m'_b,\quad\quad c_{AB}=c'_{AB},
\ee
and
\be\label{nAtrans2}
n_A={n'}_A - D_A (T {M'}_b ) - 2 {M'}_b D_A T,
\ee
where we have used the expression of $C_{AB}$ in \eqref{CAB} to verify that $D^{\bz'} D^{\bz'} {C'}_{\bz' \bz'} - D^{z'} D^{z'} {C'}_{z' z'}=0$ when deriving \eqref{nAtrans2}. The obtained transformation laws are consistent with the finite transformation results in \cite{Barnich:2016lyg,Chen:2023zpl,Flanagan:2023jio}. Clearly, the condition of preserving $n_A$ in the form of \eqref{nA} in the new coordinates system will reduce the supertranslation to one translation along time direction $u$, considering that $ M_b= - \sum{\frac{m}{(v_{\mu}n^{\mu})^{3}}}$ can not be a constant for arbitrary numbers of pointlike bodies. Even for the case of a single particle, $M_b$ is not necessary to be a constant as the boost charges of the supertranslated Schwarzschild black hole are generically non-vanishing \cite{Compere:2016hzt}. The velocity of the particle essentially is a boost of the Schwarzschild solution \cite{Bonga:2018gzr}.


\section{Discussions}

In this paper, we apply the asymptotic charge analysis to prove that the order $G$ supertranslation should be ruled out in the PM expansion which can serve as the foundation of uniquely determining the angular momentum loss in gravitational scattering. It is worthwhile to point out that ruling out the order $G$ supertranslation only means that the supertranslation parameter is fixed, but not necessary to be zero. The precise formula of the order $G$ supertranslation parameter $\bar\beta(x^A)$ can be obtained for a precise gravitational scattering, such as the gravitational Bremsstrahlung \cite{Kovacs:1978eu,Bel:1981be,Jakobsen:2021smu}. It is amusing to notice that once the order $G$ supertranslation is fixed, the complete order $G$ metric is fixed in the Bondi's framework. The key point of our prescription is that we use part of the higher order solutions in the expansion of $G$ to fix the lower order supertranslation. It is interesting to check if such an idea can be valid at even higher orders in $G$ to fix the complete supertranslation freedom in the PM expansion.

As an ending remark, we will comment on the physical meaning of fixing the order $G$ supertranslation. The supertranslation dependence in the angular momentum is nothing but the fact that the angular momentum is observer dependent. This is a well known fact even in the classical mechanics. Hence, the results of the angular momentum loss in different literature can be different depending on the reference frames. Nevertheless, they must be connected by the order $G$ supertranslation. The key issue is that if one can associate the angular momentum and its loss to the proper observer and also specify their connections to the ADM value, e.g., the intrinsic gauge and canonical gauges discussed in \cite{Veneziano:2022zwh} based on the far past conditions of the future null infinity \cite{Ashtekar:2023zul}, see also \cite{Ashtekar:1979iaf,Ashtekar:1979xeo} for earlier contributions along this direction. We conjecture that the precise form of the order $G$ supertranslation parameter $\bar\beta(x^A)$ selected by our prescription should coincide with the canonical gauge at order $G$. Here, we offer a completely different prescription that is self-contained from the null infinity analysis.


\section*{Acknowledgments}

The authors thank Zhoujian Cao, Zhengwen Liu, Xiaoning Wu, and Kai-Yu Zhang for useful discussions. The authors are grateful to Prof. Massimo Porrati for clarifying several crucial remarks. This work is supported in part by the National Natural Science Foundation of China (NSFC) under Grants No. 12475059,  No. 11935009, and No. 11905156.

\appendix

\section{Linearization at the second order}
\label{linearization}

Up to the second order of $G$, the connection is given by
\begin{multline}
    \Gamma^\alpha_{\mu\nu}={\bar\Gamma}^\alpha_{\mu\nu}+\frac{G}{2} \eta^{\alpha\beta}\left(\nb_\mu H_{\nu\beta} + \nb_\nu H_{\mu\beta} - \nb_\beta H_{\mu\nu} \right)\\
    +\frac{G^2}{2} \eta^{\alpha\beta}\left(\nb_\mu h_{\nu\beta} + \nb_\nu h_{\mu\beta}- \nb_\beta h_{\mu\nu} \right) \\
     -\frac{G^2}{2}H^{\alpha\beta}\left(\nb_{\mu}H_{\beta\nu}+\nb_{\nu}H_{\beta\mu}-\nb_{\beta}H_{\mu\nu} \right)+\cO(G^3) .
\end{multline}
We define
\be
\Theta_{\mu\nu\beta}= \nb_\mu H_{\nu\beta} + \nb_\nu H_{\mu\beta} - \nb_\beta H_{\mu\nu}  ,
\ee
and
\be
\Lambda_{\mu\nu\beta}= \nb_\mu h_{\nu\beta} + \nb_\nu h_{\mu\beta} - \nb_\beta h_{\mu\nu},
\ee
which are very useful for the computations of the curvature tensor,
\begin{align}
{R_{\mu\nu\alpha}}^\beta=&\p_\nu \Gamma^\beta_{\mu\alpha} - \p_\mu \Gamma^\beta_{\nu\alpha} + \Gamma^\tau_{\mu\alpha} \Gamma^\beta_{\nu\tau} - \Gamma^\tau_{\nu\alpha}\Gamma^\beta_{\mu\tau}\nn\\
=&{\bar R_{\mu\nu\alpha} } ^{\quad\,\,\,\beta} + \frac{G}{2}(\nb_\nu {\Theta_{\mu\alpha}}^\beta-\nb_\mu {\Theta_{\nu\alpha}}^\beta)\nn \\
&+ \frac{G^2}{2} (\nb_\nu {\Lambda_{\mu\alpha}}^\beta-\nb_\mu {\Lambda_{\nu\alpha}}^\beta) -  \frac{G^2}{2} \left[\nb_\nu (H^{\tau\beta}\Theta_{\mu\alpha\tau})-\nb_\mu (H^{\tau\beta}\Theta_{\nu\alpha\tau}) \right]\nn\\
&+\frac{G^2}{4} ({\Theta_{\mu\alpha}}^\tau {\Theta_{\nu\tau}}^\beta - {\Theta_{\nu\alpha} }^\tau {\Theta_{\mu\tau}}^\beta ) + \cO(G^3).
\end{align}
The Ricci tensor is defined from the curvature tensor as
\begin{multline}
R_{\mu\alpha}={R_{\mu\nu\alpha}}^\nu
=\bar R_{\mu\alpha}  + \frac{G}{2} \nb_\nu \left(\nb_\mu H^\nu_\alpha + \nb_\alpha H_\mu^\nu - \nb^\nu H_{\mu\alpha}\right)
- \frac{G}{2} \nb_\mu  \nb_\alpha H_\nu^\nu \\
+  \frac{G^2}{2} \nb_\nu \left(\nb_\mu h^\nu_\alpha + \nb_\alpha h_\mu^\nu - \nb^\nu h_{\mu\alpha}\right)
- \frac{G^2}{2} \nb_\mu  \nb_\alpha h_\nu^\nu \\
-  \frac{G^2}{2} \left[\nb_\nu (H^{\tau\nu}\Theta_{\mu\alpha\tau})-\nb_\mu (H^{\tau\nu}\Theta_{\nu\alpha\tau}) \right] \\
+\frac{G^2}{4} \left( {\Theta_{\mu\alpha}}^\tau {\Theta_{\nu\tau}}^\nu - {\Theta_{\nu\alpha} }^\tau {\Theta_{\mu\tau}}^\nu \right) + \cO(G^3).
\end{multline}
Then the Ricci scalar is obtained as
\begin{multline}
R=g^{\mu\alpha}R_{\mu\alpha}=\bar R - G H^{\mu\alpha}\bar R_{\mu\alpha} + G \left(\nb_{\mu}\nb_{\nu} H^{\nu\mu} -\nb^{2}H^{\mu}_{\mu} \right)\\+ G^2 (H^\mu_\nu H^{\nu \alpha} - h^{\mu\alpha})\bar R_{\mu\alpha}+G^2\left(\nb_{\mu}\nb_{\nu} h^{\nu\mu} -\nb^{2}h^{\mu}_{\mu} \right)\\
 -\frac{G^2}{2}H^{\mu\alpha}\left(2\nb_{\nu}\nb_{\mu}H^{\nu}_{\alpha}-\nb^{2}H_{\mu\alpha}-\nb_{\mu}\nb_{\alpha}H^{\nu}_{\nu} \right) \\
 -\frac{G^2}{2} \eta^{\mu\alpha }\left[\nb_\nu (H^{\tau\nu}\Theta_{\mu\alpha\tau})-\nb_\mu (H^{\tau\nu}\Theta_{\nu\alpha\tau}) \right] \\
+\frac{G^2}{4} \eta^{\mu\alpha } \left( {\Theta_{\mu\alpha}}^\tau {\Theta_{\nu\tau}}^\nu - {\Theta_{\nu\alpha} }^\tau {\Theta_{\mu\tau}}^\nu \right) + \cO(G^3).
\end{multline}
Finally, the Einstein tensor is derived as
\begin{multline}
G_{\mu\nu}=R_{\mu\nu} - \frac12 g_{\mu\nu} R=\bar{R}_{\mu\nu}-\frac{1}{2}\eta_{\mu\nu} \bar{R}\\
+\frac{G}{2}\left(\eta_{\mu\nu}H^{\alpha\beta}-H_{\mu\nu}\eta^{\alpha\beta} \right)\bar{R}_{\alpha\beta}\\
+\frac{G}{2}\left(\nb_{\tau}\nb_{\mu}H^{\tau}_{\nu}+\nb_{\tau}\nb_{\nu}H^{\tau}_{\mu}-\nb^{2}H_{\mu\nu}-\nb_{\mu}\nb_{\nu}H^{\tau}_{\tau} \right)
-\frac{G}{2}\eta_{\mu\nu}\left(\nb_{\alpha}\nb_{\beta} H^{\alpha\beta} -\nb^{2}H^{\tau}_{\tau} \right)\\
+\frac{G^2}{2}\left(H_{\mu\nu}H^{\alpha\beta}-\eta_{\mu\nu}H^{\alpha\tau}H^{\beta}_{\tau}+\eta_{\mu\nu}h^{\alpha\beta} - h_{\mu\nu} \eta^{\alpha\beta}\right)\bar{R}_{\alpha\beta} \\
+\frac{G^2}{2}\left(\nb_{\tau}\nb_{\mu}h^{\tau}_{\nu}+\nb_{\tau}\nb_{\nu}h^{\tau}_{\mu}-\nb^{2}h_{\mu\nu}-\nb_{\mu}\nb_{\nu}h^{\tau}_{\tau} \right)
-\frac{G^2}{2}\eta_{\mu\nu}\left(\nb_{\alpha}\nb_{\beta} h^{\alpha\beta} -\nb^{2}h^{\tau}_{\tau} \right)\\
+\frac{G^2}{4}\eta_{\mu\nu}\left(2\nb_{\tau}\nb_{\alpha}H^{\tau}_{\beta}-\nb^{2}H_{\alpha\beta}-\nb_{\alpha}\nb_{\beta}H^{\tau}_{\tau} \right)H^{\alpha\beta} \\
+\frac{G^2}{4}\eta_{\mu\nu} \left[\nb_\alpha (H^{\beta\alpha}{\Theta^\tau}_{\tau\beta} )- \nb_\tau (H^{\beta\alpha}{\Theta^{\tau}}_{\alpha\beta})\right] - \frac{G^2}{8}\eta_{\mu\nu} \left( {\Theta^\tau}_{\tau \beta} {\Theta_\alpha}^{\beta \alpha} - {\Theta_\alpha}^{\tau\beta} {\Theta_{\tau\beta}}^\alpha \right)\\
 -\frac{G^2}{2} H_{\mu\nu} \left(\nb_\alpha \nb_\beta H^{\alpha\beta} - \nb^2 H^\tau_\tau \right)\\
-\frac{G^2}{2}\left[\nb_\alpha (H^{\beta\alpha} \Theta _{\mu\nu\beta}) - \nb_\mu (H^{\beta\alpha} \Theta_{\alpha\nu\beta})\right] + \frac{G^2}{4} \left({\Theta_{\mu\nu}}^\alpha {\Theta_{\alpha \beta}}^\beta - {\Theta_{\mu\alpha}}^\beta {\Theta_{\nu\beta}}^\alpha \right).
\end{multline}

\section{Conservation of the effective stress tensor}
\label{conservation}

The conservation of the stress tensor is induced from the Bianchi identity of the Einstein tensor $\n_\mu G^\mu_\nu=0$. We consider the Minkowski background and the pointlike bodies source for the first order perturbative theory. Hence the zeroth order of the Einstein tensor vanishes and the first order vanishes outside the trajectory of the pointlike bodies. At the second order, the Bianchi identity outside the trajectory of the pointlike bodies leads to
\be
\nb^\mu ( E_{\mu\nu} - T^E_{\mu\nu})=0.
\ee
It is obvious that $\nb^\mu  E_{\mu\nu}=0$ which can be easily deduced from the fact that the linearized equation $E_{\mu\nu}$ at the second order is invariant under the gauge transformation
\be
h_{\mu\nu}\rightarrow h_{\mu\nu} + \nb_\mu \xi_\nu + \nb_\nu \xi_\mu.
\ee
Thus the effective stress tensor is conserved outside the trajectory of the pointlike bodies.

\section{Second order metric from the full theory}
\label{full}

The solution space of the full theory near null infinity in Bondi gauge is given by \cite{Barnich:2010eb,Flanagan:2015pxa}
\begin{multline}
ds^2=-\left[1-\frac{2 V}{\rho}+ \cO(\rho^{-2})\right]\td u^2 - 2\left[1 + \cO(\rho^{-2})\right]\td u \td \rho \\
+ \bigg[D^A \Upsilon_{AB}+ \frac{4}{3 \rho} (U_{B}+u D_{B} V)-\frac{1}{8\rho}D_B (\Upsilon_{AE}\Upsilon^{AE})  \\
 + \cO(\rho^{-2})\bigg] \td u \td x^B
 +\left[\rho^2 \gamma_{AB} + \rho \Upsilon_{AB} + \cO(1)\right]\td x^A \td x^B,
\end{multline}
where
\be
\p_u V =\frac14 D^A D^B \Lambda_{AB} -\frac18 \Lambda_{AB} \Lambda^{AB},\quad \quad \Lambda_{AB}=\p_u \Upsilon_{AB},
\ee
and
\begin{multline}
\p_u U_A=\frac14 D_A (\Upsilon_{BE} \Lambda^{BE}) - \frac14 D_B (\Upsilon^{BE} \Lambda_{EA}) + \frac12 \Upsilon_{AB} D_E \Lambda^{EB} \\
+\frac14 D^B D_A D^E \Upsilon_{EB} -\frac14 D^2 D^B \Upsilon_{BA} - u D_A (\p_u V).
\end{multline}
Suppose that $V$, $U_A$, $\Upsilon$, and $\Lambda_{AB}$ can be expanded in $G$ as
\begin{align}
&V=G M_b + G^2 m_b + \cO(G^3),\\
&U_A=G N_A + G^2 n_A + \cO(G^3),\\
&\Upsilon_{AB}=G C_{AB} + G^2 c_{AB} + \cO(G^3),\\
&\Lambda_{AB}=G N_{AB} + G^2 n_{AB}  + \cO(G^3).
\end{align}
One can recover the second order self-part solution space in the main text.

\providecommand{\href}[2]{#2}\begingroup\raggedright\endgroup

\end{document}